# Controlling the Skyrmion Density and Size for Quantized Convolutional Neural Networks

*Aijaz H. Lone[1], Arnab Ganguly[2], Hanrui Li[1], Nazek El- Atab[1], Gobind Das[2] and H. Fariborzi[1]*

*[1]Computer, Electrical, and Mathematical Science and Engineering Division, King Abdullah University of Science and Technology, Thuwal, Saudi Arabia.*

*[2]Department of Physics, Khalifa University, Abu Dhabi 12788, United Arab Emirates.*

*ABSTRACT: Skyrmion devices show energy-efficient and high-integration data storage and computing capabilities. Herein, we present the results of experimental and micromagnetic investigations of the creation and stability of magnetic skyrmions in the Ta/IrMn/CoFeB/MgO thin-film system. We investigate the magnetic-field dependence of the skyrmion density and size using polar magneto-optic Kerr effect (MOKE) microscopy supported by a micromagnetic study. The evolution of the topological charge with time under a magnetic field is investigated, and the transformation dynamics are explained. Furthermore, considering the voltage control of these skyrmion devices, we evaluate the dependence of the skyrmion size and density on the Dzyaloshinskii–Moriya interaction and the magnetic anisotropy. We furthermore propose a skyrmion-based synaptic device based on the results of the MOKE and micromagnetic investigations. We demonstrate the spin-orbit torque–controlled discrete topological resistance states with high linearity and uniformity in the device. The discrete nature of the topological resistance (weights) makes it a candidate to realize hardware implementation of weight quantization in a quantized neural network (QNN). The neural network is trained and tested on the CIFAR-10 dataset, where the devices act as synapses to achieve a recognition accuracy of ~87%, which is comparable to the result of ideal software-based methods.*

*KEYWORDS: Skyrmions, Magnetic Tunnel Junction, Magneto-Optical Kerr Effect, Micromagnetics, Topological Resistance, and Neuromorphic devices*

## I. INTRODUCTION

Magnetic skyrmions are topologically protected swirling structures obtained using chiral interactions in noncentrosymmetric magnetic compounds or thin films exhibiting broken inversion symmetry [1]. The intrinsic topological protection of the skyrmion makes them stable against external perturbations [2]. Topological protection means that skyrmions have a characteristic topological integer that cannot change by the continuous deformation of the field [3]. The Dzyaloshinskii–Moriya interaction (DMI), as a chiral antisymmetric exchange interaction responsible for the formation of magnetic skyrmions, is based on the strong spin-orbit coupling at the heavy metal/ferromagnetic (HM/FM) interface with broken-inversion symmetry [4]. Magnetic skyrmions emerge from the interaction between different energy terms. The exchange coupling and anisotropy terms support the parallel alignment of spins, whereas the DMI and dipolar energy terms prefer the noncollinear alignment of spins [5]. In asymmetric FM multilayer systems, such as Pt/Co/Ta [6] and Pt/CoFeB/MgO [7], the DMI is facilitated by the high interfacial spin-orbit coupling induced by symmetry breaking. Because the DMI and anisotropy terms are material property- and geometry-dependent, a combination of different HM/FM structures has been investigated to stabilize skyrmions and define a specific chirality [8]. Spintronic devices based on magnetic skyrmions exhibit increased density and promote energy-efficient data storage because of their nanometric size and topological protection [9]. Magnetic skyrmions can be driven by extremely low depinning current densities [10], and they can be scaled down to 1 nm [11]. These properties indicate that they can be potentially applied in data storage and computing [12]. In particular, skyrmion devices exhibit excellent potential for application in unconventional computing techniques such as neuromorphic computing [13] and reversible computing [14]. Neuromorphic computing is inspired by the performance and energy efficiency of the brain [15]. Furthermore, it employs neuromimetic devices, emulating neurons, for computing tasks. Synapses store information in terms of weight. Spintronic devices, particularly magnetic tunnel junctions (MTJs), have attracted considerable interest in neuromorphic computing [16][17][18][19]. Recently, multiple neuromorphic computing systems coupled with MTJs based on skyrmions, such as skyrmion neurons [20][15][21] and skyrmion synapses [13][22][23], have been proposed. Furthermore, the control of spintronic devices using an electric field has attracted considerable attention for memory and logic applications because it is an efficient approach for improving the data-storage density [24][25] and reducing the switching energy [26]. The important challenges encountered in the application of skyrmions in storage and computing (both conventional and unconventional computing) are the controlled motion and readability of skyrmions [27].

Quantization of neural network is an effective method for model compression and acceleration. To compress the network, all weights of the model with a high-bit floating-point number are quantized into a low-bit integer or fixed floating number, which significantly reduces the amount of the required memory size and computational cost [28]. Emerging memory devices could emulate functions of biological synapses to realize in-memory computing, which paved the way for the quantized weight implementation in the neuromorphic computing system [29][30]. In this study, we conduct experimental and micromagnetic investigations in magnetic skyrmions in the Ta/IrMn/CoFeB/MgO thin film system for realization of neuromorphic device. The dependence of the skyrmion density and diameter on the magnetic field is studied by polar magneto-optic Kerr effect (MOKE) microscopy supported with the micromagnetic simulations. The evolution of the topological charge with time under the magnetic field is investigated, and the transformation dynamics from the labyrinth domains to Neel skyrmions is explained. Furthermore, we evaluate the dependence of the skyrmion density and size on the DMI and the magnetic anisotropy to realize voltage control skyrmion devices. Based on these results of the MOKE and micromagnetic investigations, we propose a skyrmion-based synaptic. We demonstrate the spin-orbit torque (SOT)-controlled skyrmion device and its constituent discrete topological resistance states. Considering the discrete nature of the topological resistance (weights), we demonstrate the neuromorphic implementation based on the quantized convolutional neural network (QCNN) architecture. The NN is trained and tested on the CIFAR-10 dataset and the devices acting as synapses achieve a recognition accuracy of ~90%, which is comparable to the accuracy of ideal software-based methods.

## II. Methods

### Fabrication and Characterization

As shown in Fig. 1, a multilayer thin film of Ta (5 nm)/IrMn (5 nm)/CoFeB (1.04 nm)/MgO (2 nm)/Ta$_2$ (nm) was deposited on thermally oxidized Si substrates by Singulus DC/RF magnetron sputtering. In this sample, the thickness of CoFeB is a curtailment parameter that provides suitable anisotropy for creating high-density skyrmions. The sputtering conditions were carefully optimized to achieve perpendicular magnetic anisotropy (PMA). Then, the sample was subjected to a post-deposition annealing treatment at 250°C for 30 min to enhance the PMA. The investigations were performed using MOKE microscopy in the polar configuration. Differential Kerr imaging was performed to observe the magnetic domains and eliminate the contribution of any nonmagnetic intensity. The square pulses of the MF were simultaneously applied both in-plane and out-of-plane of the sample using two independent electromagnets. The sample exhibited a labyrinth domain structure without an MF. First, using a sufficiently large out-of-plane field, magnetization was saturated in one perpendicular direction. A reference polar-MOKE image was captured in this state. Next, the out-of-plane MF strength ($H_z$) was decreased to the desired value while the in-plane field with strength $H_x$ was applied, as required. Subsequently, a second polar-MOKE image was captured in this state. The magnetic image of the final state was obtained by the differential image relative to the reference image. Fig. 1(b) shows the typical double-loop magnetic hysteresis characteristic of the multidomain in the system, which supports the MOKE imagining results.

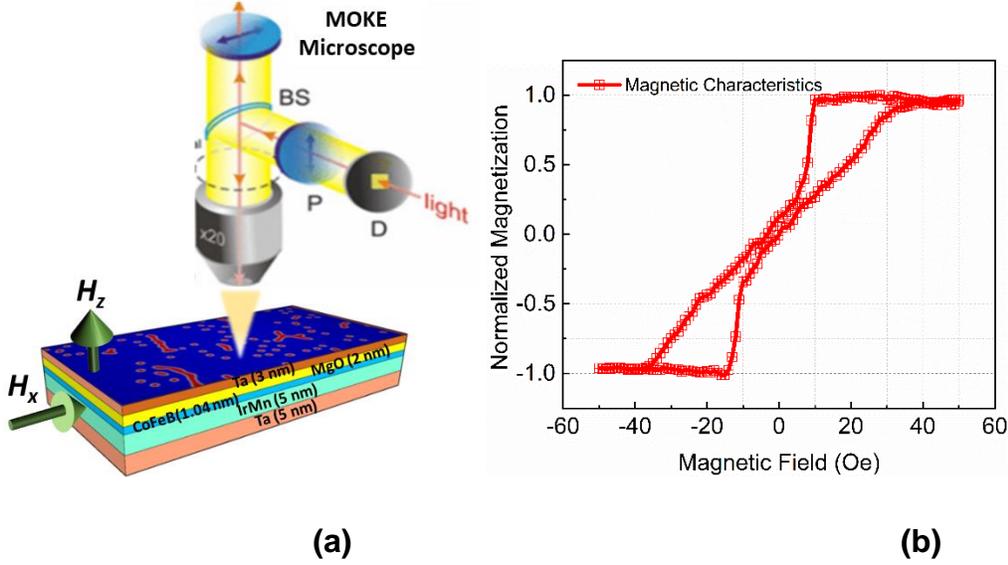

(a)            (b)

*Figure 1: Schematic of the fabricated sample for MOKE and vibrating sample magnetometry (VSM) characterizations*

MICROMAGNETICS

Magnetic skyrmions are described using their topological or skyrmion number (Q), calculated as follows [31]:

$$Q = \frac{1}{4\pi} \int \int \boldsymbol{m} \cdot \left( \frac{\partial \boldsymbol{m}}{\partial x} \times \frac{\partial \boldsymbol{m}}{\partial y} \right) dx dy. \qquad (1)$$

The spins projected on the xy plane and the normalized magnetization vector (***m***) can be determined by the radial function ($\theta$), azimuthal angle $\varphi$, vorticity ($Q_v$), and helicity ($Q_h$):

$$m(r) = [\sin(\theta)\cos(Q_v\varphi + Q_h), \sin(\theta)\sin(Q_v\varphi + Q_h), \cos(\theta)]. \qquad (2)$$

The vorticity is related to the skyrmion number via the following expression [12]:

$$Q = \frac{Q_v}{2} \left[ \lim_{r \to \infty} \cos(\theta(r)) - \cos(\theta(0)) \right]. \qquad (3)$$

Micromagnetic simulations were performed using MuMax having the Landau–Lipschitz–Gilbert (LLG) equation as the basic magnetization dynamics computing unit [32]. The LLG equation describes the magnetization evolution as follows:

$$\frac{d\hat{m}}{dt} = \frac{-\gamma}{1+\alpha^2}[\boldsymbol{m} \times \boldsymbol{H}_{eff} + \boldsymbol{m} \times (\boldsymbol{m} \times \boldsymbol{H}_{eff})], \qquad (4)$$

$\gamma$ is the gyromagnetic ratio, $\alpha$ is the Gilbert damping coefficient, and

The total magnetic energy of the free layer includes the exchange, Zeeman, uniaxial anisotropy, demagnetization, and DMI energies [33][34]:

$$E(\boldsymbol{m}) = \int_V [A(\nabla \boldsymbol{m})^2 - \mu_0 \boldsymbol{m}.\boldsymbol{H}_{ext} - \frac{\mu_0}{2}\boldsymbol{m}.\boldsymbol{H}_d - K_u(\hat{u}.\boldsymbol{m}) + \boldsymbol{\varepsilon}_{DM}]dv, \qquad (5)$$

where A is the exchange stiffness, $\mu_0$ is the permeability, $K_u$ is the anisotropy energy density, $H_d$ is the demagnetization field, and $H_{ext}$ is the external field. Moreover, the DMI energy density is computed as follows:

$$\varepsilon_{DM} = D[m_z(\nabla.\boldsymbol{m}) - (\boldsymbol{m}.\nabla).\boldsymbol{m}]. \qquad (6)$$

$$\boldsymbol{H}_{eff} = \frac{-1}{\mu_0 M_S}\frac{\delta E}{\delta \boldsymbol{m}} \qquad (7)$$

is the effective magnetic field around which the magnetization vector precesses.

Then, the SOT is introduced as a custom field term into MuMax [35]:

$$\boldsymbol{\tau}_{SOT} = -\frac{\gamma}{1+\alpha^2}a_J[(1+\xi\alpha)\boldsymbol{m}\times(\boldsymbol{m}\times\boldsymbol{p}) + (\xi-\alpha)(\boldsymbol{m}\times\boldsymbol{p})], \quad (8)$$

$$a_J = \left|\frac{\hbar}{2M_S e\mu_0}\frac{\theta_{SH}j}{d}\right| \quad \text{and} \quad \boldsymbol{p} = sign(\theta_{SH})\boldsymbol{j}\times\boldsymbol{n},$$

where $\theta_{SH}$ is the spin Hall coefficient of the material, $j$ is the current density, and $d$ is the free-layer thickness. The synapse resistance and neuron output voltage are computed using the nonequilibrium Green's function (NEGF) formalism. We then consider the magnetization profile of the free layer and feed it to the NEGF model, which computes the resistance of the device, as follows [36][37]:

$$R_{syn} = \frac{V_{syn}}{I_{syn}}. \qquad (9)$$

Subsequently, the MTJ read current is computed as follows:

$$I_{syn} = trace\left\{\sum_{k_t} C_\sigma \frac{i}{\hbar}\begin{Bmatrix} H_{k,k+1}\, G^n_{k+1,k} \\ -G^n_{k,k+1} H_{k+1,k} \end{Bmatrix}\right\}, \quad (10)$$

where $H_k$ is the $k$th lattice site in the device Hamiltonian, and $G^n_k$ is the electron correlation at the $k$th site, which yields the electron density.

## III. RESULTS AND DISCUSSION

In this study, skyrmions were created and annihilated in a multilayer thin film using a combination of in-plane and out-of-plane MFs. The sample and experimental conditions were optimized to achieve a relatively high skyrmion density at a low field amplitude. We experimentally obtained a maximum skyrmion density of $350 \times 10^3$ /mm$^2$ for the resultant MFs with $H_z$ and $H_x$ values of 20 Oe and 35 Oe, respectively. As shown in Fig. 2(a), we experimentally observed the evolution from the labyrinth domains into skyrmions under a constant in-plane MF ($H_X$ = 2 mT) and an out-of-plane MF ($H_z$ = 0–2.4 mT), as reported in the top-right corner of each figure. The white and black contrasts correspond to the ↑ and ↓ domains, respectively. We then observed that the large labyrinth domains start splitting into small domain structures, and skyrmions start emerging as $H_z$ increases from 0 to 1.2 mT. At $H_z$ = 2 mT, all the labyrinth domains disappear and the skyrmion density reaches the maximum value. Then, any additional increase in the field strength decreases the skyrmion density because of skyrmion annihilation. The complete annihilation of the skyrmions occurs when $H_z$ is ~2.6 mT. The magnetization reversal mechanism is analytically explained using micromagnetic simulations. Here, we employ the finite difference method to solve the LLG equation to examine the spin dynamics in a system similar to that used in the experiments. We consider a 1024 × 512 × 1-nm$^3$ free layer discretized as a mesh with 1 × 1 × 1-nm$^3$ cells. The magnetic parameters used for the simulations are as follows: anisotropy, K = 1.1 × 10$^6$ J/m$^3$; exchange constant, A = 1 × 10$^{11}$ J/m; and saturation magnetization, M$_s$ = 800 A/m. Fig. 2(b) shows that the micromagnetic simulation results are consistent with the experimental results. At $H_z$ = 2 mT, we observe the splitting of large labyrinth domains into small domain structures, followed by the gradual stabilization of skyrmions; accordingly, an increased skyrmion density is observed at $H_z$ = 10 mT. An important observation from these simulations is that the magnetic domains split, rotate counterclockwise, and shrink until the skyrmions are stabilized. However, during this process, the already stabilized skyrmions are robust, and except for small translation motions, we observe no change in their size. Once only skyrmions are present in the magnetic free layer, any additional increase in the field strength will cause the skyrmion size to shrink. However, at this point, the skyrmions are less responsive to the MF, depicting their topological stability. The experimental field dependence of the skyrmion density is summarized in a 3D color plot in Fig. 2(c). The horizontal axes represent $H_x$ and $H_z$, and the height and the color represent the corresponding skyrmion densities. The skyrmion density peaks at $H_z = 2$ mT, independent of $H_x$ (black dashed line). However, the skyrmion density is not symmetric with respect to the $H_z = 2$ mT line and is larger for higher values of $H_z$. The red and blue arrows in Fig. 2(c) show the asymmetry of the skyrmion density line with respect to $H_z = 2\ mT$. The degree of asymmetry increases with $H_x$. This result demonstrates that the evolution of skyrmions involves two different phases. On the left side of $H_z = 2$ mT, the skyrmion creation rate is higher than the annihilation rate, indicating the net creation of skyrmion. On the right side, net annihilation occurs. At high $H_x$ values, a larger field range is required for skyrmion annihilation than that for creation. This observation can be explained by the increase in the skyrmion size with $H_x$ and the topological protection of skyrmions. Therefore, for annihilation, large fields are required to decrease the critical diameter of skyrmions. Thus, at high $H_x$ values, additional energy is required for the annihilation of skyrmions. This result demonstrates that the in-plane field promotes the stability of the skyrmion spin structures; as such, their annihilation occurs at high chiral energies. This can be observed in Figs. 2(b) and 2(e), where the skyrmions are created under fields with $H_z = 40$ mT; however, the annihilation occurs at ~100 mT. When comparing to MOKE results as shown in Fig. 2(d), we observe micromagnetic simulations agreeing very well qualitatively (see Fig. (e)). Moreover, the skyrmion density increases with $H_x$, and a relatively broad window of $H_z$ is observed, indicating the presence of skyrmions. When an in-plane field is applied to the labyrinth domains, the domains align along the field direction and their widths decrease. The aligned domains increase the efficiency of the skyrmion creation compared to the case with misaligned labyrinth domains. The experimental results qualitatively agree with simulation results (Fig. 2(e)). The missing quantitative aspect is justified by the difference in the sample sizes and time differences in application of magnetic field in the experiment and simulation. The simulation time considered in MuMax was in range of ns while the magnetic field was applied for 100ms. In the experiment, the stack area was 100 mm$^2$, whereas in the simulation, considering the computational viability, the MuMax stack size was set to 1.024 μm × 512 nm.

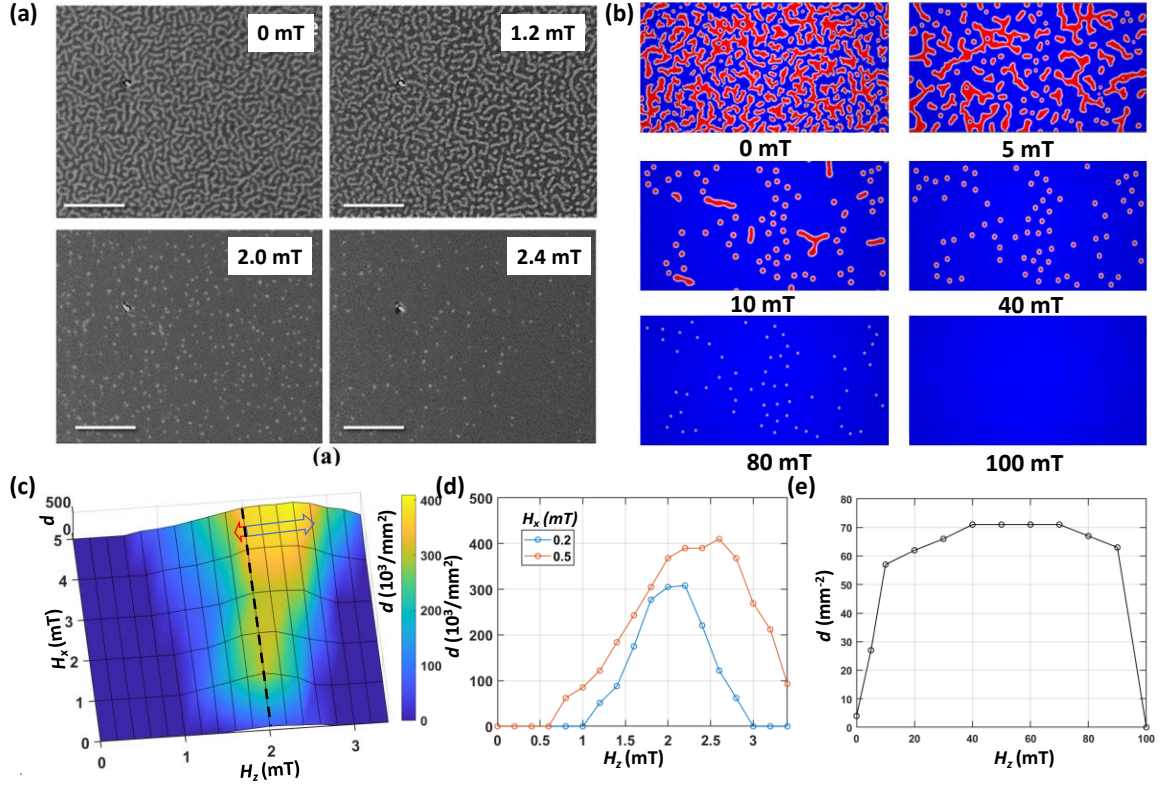

*Figure 2. (a) MOKE captures of the sample showing the MF dependence of the skyrmion density. (b) Micromagnetic simulation results corresponding to the experimental results. (c) Skyrmion-density tuning by applying in-plane and perpendicular MFs. (d) Skyrmion density as function of Hz (MOKE) (e) Skyrmion density versus $H_z$ showing an increasing trend, followed by a saturated constant (wide window) and, subsequently, annihilation (Micromagnetics)*

Fig. 3(a) shows the dynamics of the skyrmion stabilization. We observe that the highly asymmetrical domains at t = 3 ns rotate counterclockwise while gradually shrinking to additional symmetrical textures; thereafter, they are transformed into topologically protected skyrmions at t = 14.2 ns. The rotation stops when the domain transforms into a highly symmetrical skyrmion. Thus, the magnetic energy at the beginning is dissipated in three degrees of freedom (DOFs): rotation, translation, and shrinkage. Gradually, the DOFs are restricted; in particular, the rotation DOF completely vanishes with time because of the formation of symmetrical magnetic textures (skyrmions). In Figs. 3(b) and (c), we demonstrate the different torques acting on the domains. For the asymmetrical domains, the unbalanced torques due to the in-plane component ($H_x$) and the perpendicular component ($H_z$) induce domain rotation and domain shrinkage, respectively. However, for skyrmions, the torques are rotation torques, and they are balanced because of the symmetrical magnetic textures; thus, the rotation DOF is eliminated. Consequently, the skyrmion size gradually decreases because of the increase in the shrinking energy.

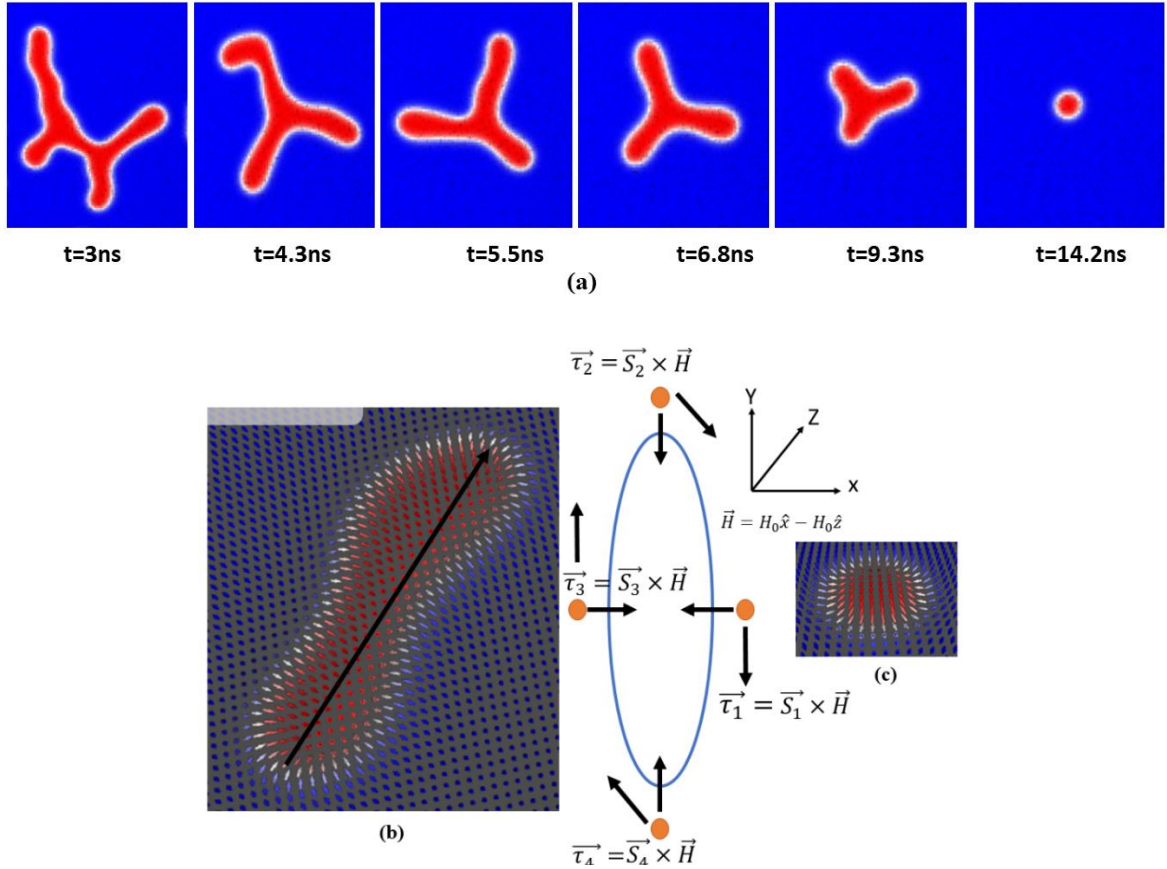

Fig. 3 (a) Skyrmion-stabilization dynamics showing the counterclockwise rotation of the labyrinth domains and the gradual shrinking to yield skyrmions. (b) The unbalanced torque caused by the field in the asymmetrical domains causes rotation due to $H_x$ and gradual shrinking due to $H_z$. (c) For the skyrmion, the symmetrical texture balances the net torque, and hence no rotation occurs.

In Figs. 4 (a1–a3), we demonstrate the evolution of the total topological charge under the MF at different times. The topological charge is initially −10, which indicates that skyrmions with $Q = -1$ dominate the overall free-layer magnetic texture. However, under the field with $H_Z = -30$ mT, the topological charge switches to +55, and within a fraction of a nanosecond, Q settles at around 40. On increasing the MF, the $Q = +1$ skyrmion density increases, as shown in Figs. 4(a2) (d–f). The maximum topological charge reaches 50 when all the $Q = -1$ skyrmions are annihilated, and we observe only Q = +1 skyrmions, as shown in (f). The Q = +1 skyrmions are stable from 40 to 80 mT, as shown in Fig. 2(e), and the corresponding topological charge is fixed at 50 in this range, as shown in Fig. 4(b).

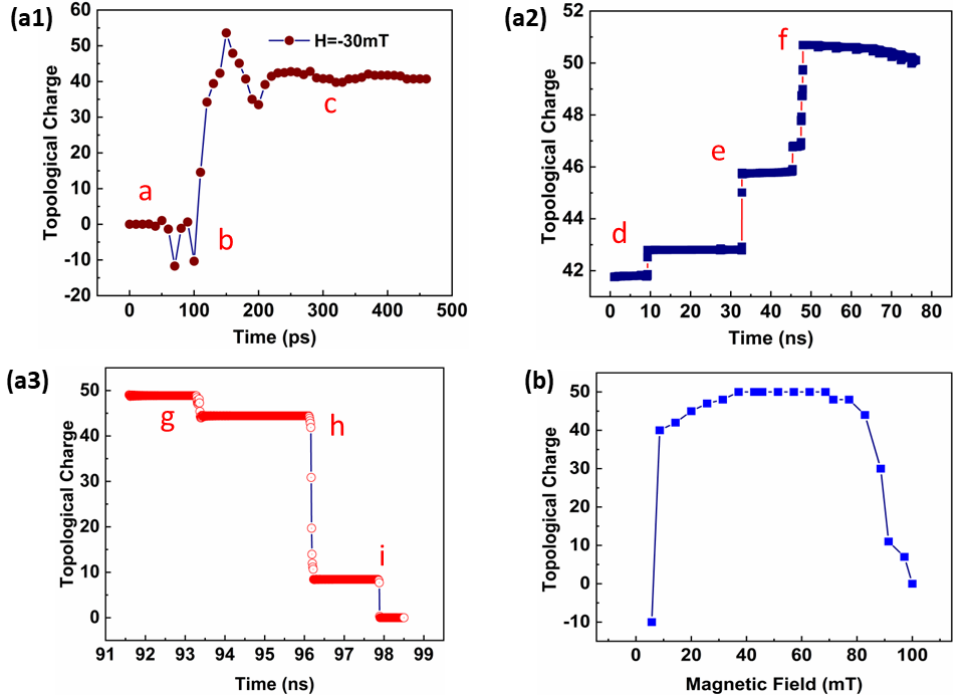

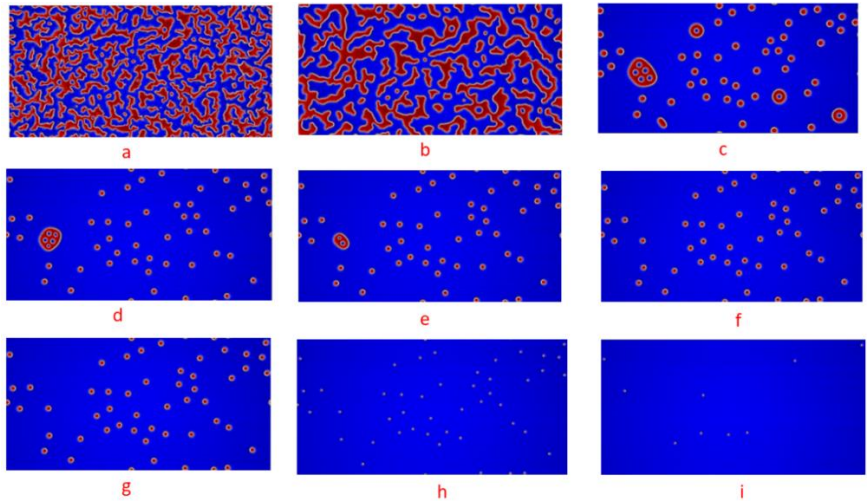

*Fig. 4 (a1–a3) Time evolution of the topological charge (Q) under the MF. (b) Topological charge vs MF. (c) Magnetic texture corresponding to different times and fields.*

As we increase the field, the skyrmion size decreases and we start observing the annihilation of Q = +1 skyrmions and the decrease in the topological charge. The topological charge reaches 0 around $H_Z$ = 100 mT.

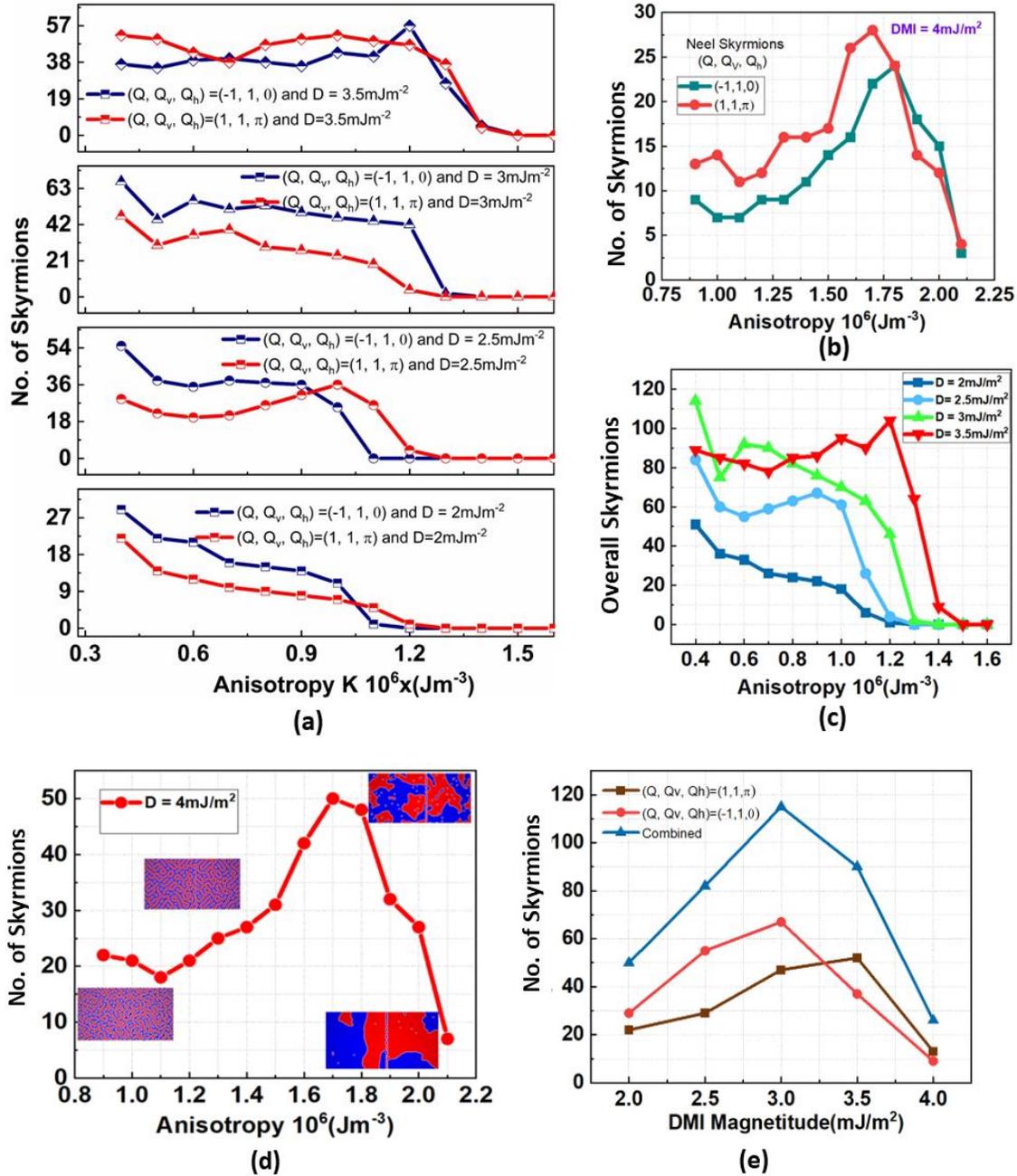

*Figure 5. (a) Anisotropy dependence of the skyrmion density for different DMI magnitude; the plots show the behavior of two types of skyrmions with $(Q, Q_V, Q_h) = (−1, 1, 0)$ and $(1, 1, 0)$. (a–e) DMI magnitude = 2, 2.5, 3, 3.5, and 4 mJ/m², respectively. (b) DMI = 4mJm⁻² (c) Combined skyrmion density versus anisotropy while increasing the DMI magnitude. (d) DMI = 4mJm⁻²  (e) The control of the skyrmion density with increasing the DMI*

In Figs. 5(a), we show the skyrmion density dependence on the DMI and anisotropy, for skyrmion with indices Q=(-1,1,0) and (1, 1, π)The simulations were carried out for DMI in range (2mJm-2 to 4mJm-2) and anisotropy starting from $K = 0.5 \times 10^6$ J/m$^3$ to $K = 1.6 \times 10^6$ J/m$^3$. In all simulations, we observe the presence of two types of skyrmions with different winding numbers (charges, Q), vorticities ($Q_v$), and helicities ($Q_h$) in the same FM thin film. The magnetization texture splits into large domains, and these domains, in turn, split into small labyrinth domains and skyrmions. Depending on the background magnetization of these large domains, the skyrmions have (Q, $Q_v$, $Q_h$) either as (−1, 1, 0) or (1, 1, π). Here, we independently consider the impact of the DMI and anisotropy on the size and density of these two types of skyrmions. At the DMI magnitude (DMI) of $2 \times 10^{-3}$ J/m$^2$, the density of skyrmions with attributes (1, 1, π) gradually decreases smoothly with an increase in the anisotropy and undergoes complete annihilation at $K = 1.3 \times 10^6$ J/m$^3$. However, the skyrmions with attributes (−1, 1, 0) undergo abrupt annihilation at ~$K = 1.1 \times 10^6$ J/m$^3$, indicating their lower stability than those of the other skyrmions. As DMI increases to $2.5 \times 10^{-3}$ J/m$^2$, the skyrmion-density behavior compared to the anisotropy starts changing, the number of skyrmions with attributes (1, 1, π) increases from 29 to the local minimum of 20, and a peak value of 35 is observed at $K = 1 \times 10^6$ J/m$^3$, followed by sharp annihilation at $K = 1.25 \times 10^6$ J/m$^3$. The density of the other type of skyrmion remains almost constant until $K = 1 \times 10^6$ J/m$^3$, followed by an abrupt decay at $K = 1.1 \times 10^6$ J/m$^3$. At DMI $= 3 \times 10^{-3}$ J/m$^2$, the skyrmion (1, 1, π)'s density gradually decreases with a few oscillations, whereas that of the skyrmion with the (−1, 1, 0) attribute remains constant until $K = 1.2 \times 10^6$ J/m$^3$, after which a sharp annihilation is observed. With an additional increase in DMI, the skyrmion density demonstrates a more stable behavior with an increase in the anisotropy, followed by an abrupt annihilation. For both types of skyrmions, at DMI $= 4 \times 10^{-3}$ J/m$^2$, the skyrmion density decreases at low anisotropies and increases with the anisotropy to a peak value as shown in Fig. 5(b). Thus, a normal trend of the skyrmion density decreasing with the anisotropy is observed, as in previous cases. We plot the combined skyrmion density in Fig. 5(c) and Fig. 5(d), these results demonstrate an improved image of the skyrmion-stability dependence on the DMI and anisotropy. If the DMI of a system is low, skyrmions can exist only at low anisotropies; however, the stability increases with the DMI magnitude, and the skyrmions can exist in a range of anisotropies. At high DMI values, the reverse behavior is observed, and the skyrmion density attains the maximum value at a high anisotropy. The skyrmion density simultaneous dependence on DMI is shown in Fig. 5(e), we observe the maximum skyrmion density around DMI = $3 \times 10^{-3}$ J/m². As both DMI and anisotropy depend upon the FM thickness and spin orbit interaction. This study provides additional insights into the optimization of material and geometrical properties, particularly the thickness of FM thin films, for the stabilization of skyrmions; for example, for a stable racetrack memory, conventional and unconventional logics depend on the SOT-driven motion of skyrmions.

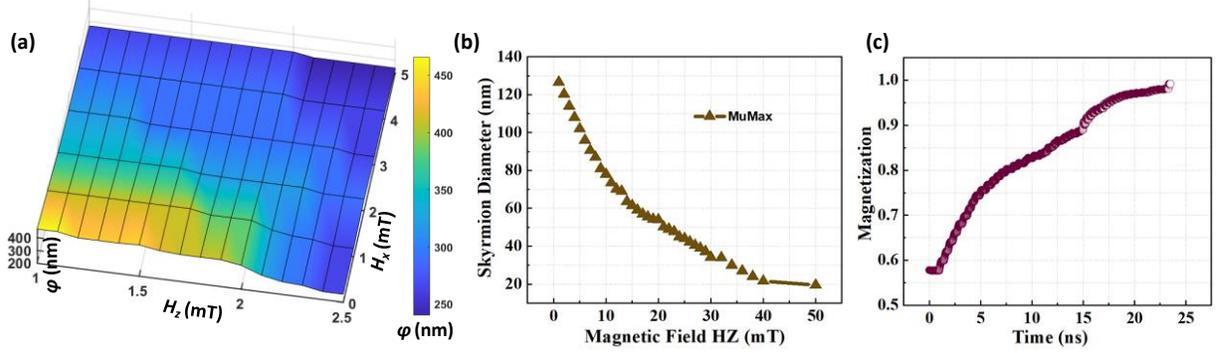

*Figure 6. (a) Skyrmion size decreasing with increasing $H_x$ and $H_z$ (MOKE). (b) The micromagnetic simulation captures a similar trend. (c) Magnetization behavior with time shows the gradual decrease in the responsiveness of the skyrmion to the field*

The optimum DMI magnitude and anisotropy ranges are 3.1–3.5 mJ/m² and $0.5–1.2 \times 10^6$ J/m³, respectively, as the skyrmion density remains almost constant in these regions, which is important for realizing a reliable memory/logic operation. However, low DMI values (2–2.5 mJ/m²) appear to be ideal for voltage-controlled operations. Considering the behavior in Figs. 5(a) and Fig. 5(c) for D < 2.5 mJ/m², we observe a smooth variation in the skyrmion density with the anisotropy. The dependence of anisotropy on voltage has been demonstrated in [38]:

$$K_S(V) = K_S(0) - \frac{\xi E}{t_{FL}}, \qquad (11)$$

where $K_S(V)$ is the anisotropy at voltage $V$, $E$ is the electric field across the oxide, $\xi$ is the voltage-controlled magnetic anisotropy (VCMA) coefficient, and $t_{FL}$ is the free-layer thickness. In Fig. 6(a), the MOKE results demonstrate that the skyrmion size decreases with an increase in $H_x$ and $H_z$. The skyrmion size is the maximum at $H_z$ = 1 Mt and $H_x$ = 0; the size decreases with an increase in both $H_x$ and $H_z$. Fig. 5(b) shows how the skyrmion responds to the magnetic field and that the size of the skyrmion sharply decreases. However, as the size decreases, the intraskyrmion forces increases, decreasing the responsiveness of the skyrmion to the MF. Consequently, the skyrmion size decreases more gradually than the case with the large skyrmion. Fig. 6(c) shows the free-layer magnetization having a single skyrmion under the magnetic field. As the skyrmion size decreases, the magnetization increases; however, with time, saturation is realized because of the increase in the strength of the intraskyrmion forces, and the magnetic field finally overcomes the topological barrier. Thus, the skyrmion is annihilated under a strong field.

Fig. 6(a) shows the change in the magnetization texture of the free layer with an increase in the anisotropy; the skyrmion density and size decrease with an increase in the anisotropy. For our device simulations, we consider $\xi = 130$ (fJ(Vm) $- 135$] and $t_F = 1\ nm$. Therefore, we used the VCMA to control the skyrmion density, which controls the synaptic conductance and demonstrates its neuromimetic behavior. Fig. 6(b) shows the MuMax simulations, showing the variations in the skyrmion size with the anisotropy. The micromagnetic simulation exhibits a similar trend to that observed in the experiment: at the beginning, the skyrmion size decreases rapidly; however, the skyrmion becomes more stable and unresponsive to the magnetic field as the size decreases because of the intraskyrmion forces (topological barrier). Fig. 6(c) shows the variation in the skyrmion size with an increase in the anisotropy at a constant DMI value (3 mJ/m$^2$). The skyrmion diameter linearly decreases in the anisotropy range of 0.58–0.64 × 10$^6$ J/m$^3$ and at both extreme ends of the anisotropy range. The magnetization behavior with time plot shows a gradual decrease in the responsiveness of the skyrmion to the field as well as the final annihilation under a considerably strong field.

Thus, we can exploit the linear-region response for the synapses that act as linear weights; furthermore, the overall skyrmion behavior has considerable relevance to the sigmoid neuron behavior in artificial neural networks. We express this behavior in terms of a fitting model, as a modified sigmoid function:

$$R = \frac{\beta D}{1 + e^{c_1(K - c_2)} + e^{c_1(K + c_2)}}. \qquad (12)$$

From the equation, the critical condition for a skyrmion to exist is derived after expanding the equation to the first order in anisotropy K. We obtained the radius dependence on anisotropy K for a fixed DMI value of 3 mJ/m$^2$, as follows:

$$R = \frac{\beta D}{3 + 2C_1 K}. \qquad (13)$$

$\beta$, $c_1$, and $c_2$ are the fitting coefficients ($1.03 \times 10^5$ m$^3$/J, $5 \times 10^{-5}$ m$^3$/J, and $6.1 \times 10^5$ J/m$^3$, respectively). The simulation results agree with the fitting model results.

These results open another possibility for the realization of voltage-controlled neuromorphic skyrmion devices. On application of the voltage, the anisotropy decreases or increases. Thus, a corresponding variation in the skyrmion size can be obtained.

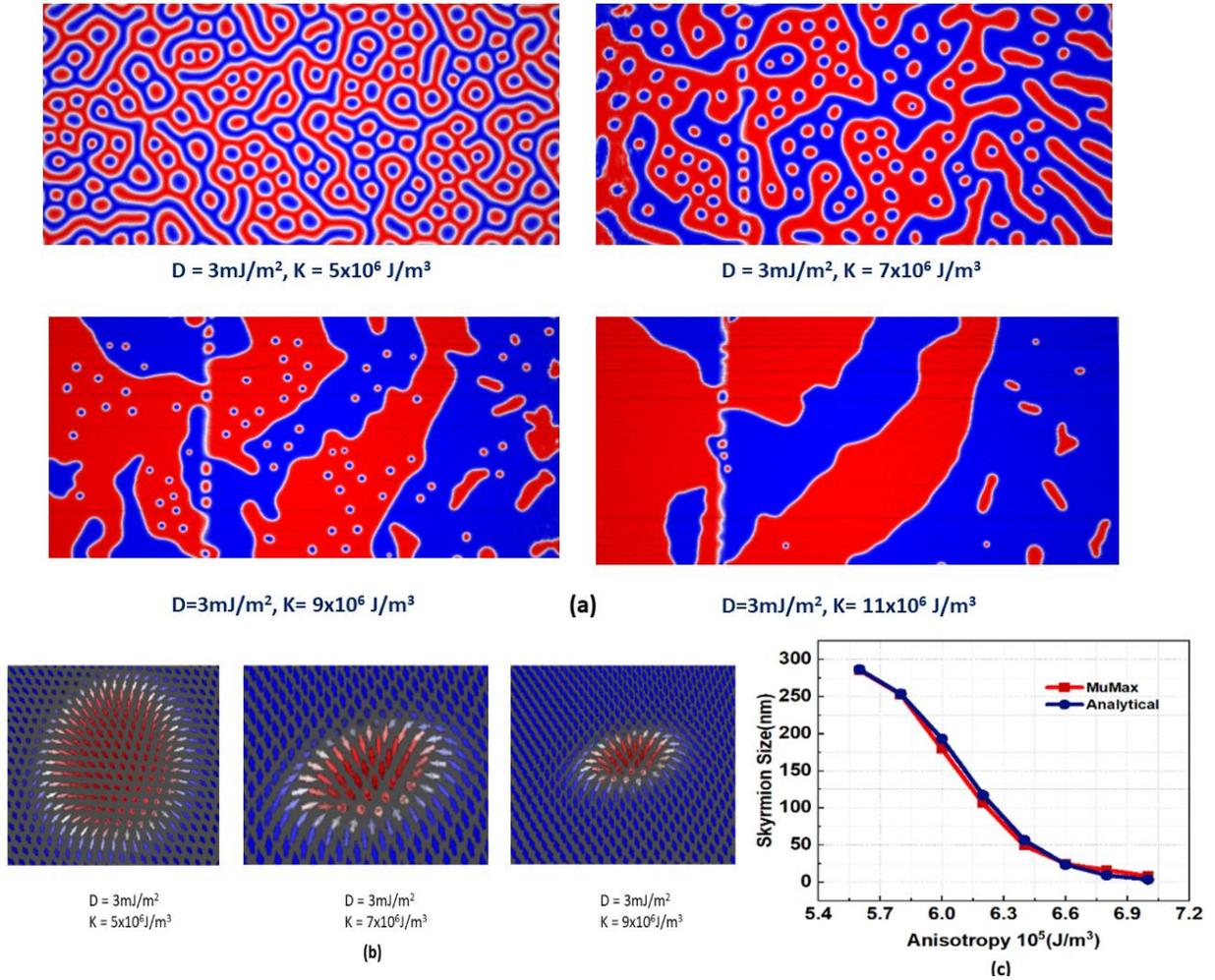

*Figure 7. (a) The CoFeB (1 nm) layer magnetization texture evolution with anisotropy showing that the skyrmion density remains almost constant at 5–7 × 10$^5$ J/m$^3$ and decreases with a further increase in K. The skyrmions are annihilated at ~K = 11 × 10$^6$ J/m$^3$. (b) The skyrmion size at a constant DMI value decreases with an increase in the anisotropy. (c) The skyrmion-size dependence on the anisotropy at a constant DMI of 3 mJ/m$^2$ agrees with the fitted analytical model. The model can be used as a neuron thresholding function in spiking and artificial neural networks.*

The results based on the skyrmion density and size manipulation using the magnetic field, DMI, and anisotropy (voltage) terms provide in-depth insight into the optimization of materials' parameters, stack geometry, and switching techniques for tuning the skyrmion density and diameter for memory and logic applications.

## MODULATING THE SKYRMION DENSITY AND SIZE FOR A QUANTIZED CONVOLUTIONAL NEURAL NETWORK

Motivated by the skyrmion density, topological charge and skyrmion-size modulation discussed thus far, in Fig. 7(a), we propose a skyrmion-based memristive device, where the skyrmion topological resistance increases/decreases when a skyrmion is moved into/out of the active region. The current is applied across T-1 and T-2, which drives the skyrmions from pre-synapse to main-synapse. The topological Hall resistance due to the skyrmions is expressed as follows[40]:

$$B_Z^e = \frac{\Phi_Z^e}{A} = -\frac{h}{eA}\iint \frac{1}{4\pi}\bm{m}.\left(\frac{\partial \bm{m}}{\partial x}\times\frac{\partial \bm{m}}{\partial y}\right)dxdy,$$

$$\rho_{xy}^T = PR_o\left|\frac{h}{e}\right|\frac{1}{A}.$$

Here, $P$ is the spin polarization of the conduction electrons, $R_o$ is the normal Hall coefficient, $h$ is Planck's constant, e is the electron charge, A is the area of the cross-overlap, and $\frac{h}{e}$ is the flux quantum. The topological resistivity change is measured across T3 and T-4. Following the results from[2], the conductivity contribution by one skyrmion is 22 $n\Omega$cm. Therefore, in the proposed skyrmion synapse, the topological resistance (RTHE) across XY is expected to increase by 22 $n\Omega$ on adding each skyrmion to the synapse region. We create a discrete set of skyrmions in the pre-synapse region, as shown in Fig. 7(b). Thereafter, using SOT current pulses, the skyrmions are driven into the synapse region, and the corresponding topological resistivity is reflected in RAH across XY. The skyrmions move at 80 m/s; thus, roughly each skyrmion takes time = sky-position(initial)/velocity to reach the main-synapse region. We observe that the first skyrmion located at −50 nm from the center arrives in ~0.6 ns, and a step in the topological charge is detected. Likewise, in the constant current pulse, other skyrmions move into the synapse region, as shown in Fig. 7(b), and we observe discrete steps, as shown in Fig. 7(c). For the 8-skyrmions, eight discrete steps are detected. This results in discrete topological resistivity, as shown in Fig. 7(d). For the 16-skyrmions, the resistivity increases in 16 discrete steps on the application of current pulses. As shown in Fig. 7(b), the current moves from +x to −x, and the skyrmions move from −x to +x. Assuming that the background magnetization is not impacted by the current but by the skyrmion motion, then during the period before the arrival of the next skyrmion, the resistivity remains constant. Thus, we observe that the topological resistivity increases in discrete steps, as shown in Fig. 7(d) (Potentiation). After 16 pulses, the current direction is reversed, the skyrmions move from +x to −x, and the topological charge in the main synapse decreases, which reduces the topological resistivity. This is referred to as synaptic depression. This is justified in Fig. 7(d), where it is shown that the topological charge in the synapse region increases from 0 to 15, and this is referred to as synaptic potentiation. To induce the synaptic depression, the current direction is reversed and the skyrmions are removed from the active region into the pre-synapse region. Note that any number of discrete states, such as 8 (3bit), 16 (4bit), 32 (5bit), and 64 (6bit) discrete states, can be realized by creating the corresponding skyrmions in the pre-synapse region. However, with increase in the skyrmion density, the skyrmion–skyrmion repulsion and skyrmion-hall effect begin to distort the skyrmion size, as shown in Fig. 7(b), although no impact on the topological charge is observed. This indicates that the topological resistivity–based skyrmion synapse tends to be more stable and noise resilient. In Fig. 7(e), the magneto-tunnel resistance corresponding to the discrete skyrmions in the synapse is shown. If the device resistance is measured using a MTJ, we observe that with each skyrmion moving into or out of the synapse, the vertical tunnel resistance varies by 5.62 $\Omega$ (73 m$\Omega$.cm). In addition to the MF and SOT control, the variation in the skyrmion size with MF and VCMA discussed in this work can introduce extra novelty to skyrmion device design. In particular, the tunnel resistance exhibited by the device can be tuned by SOT and voltage control; thus, additional advanced functional devices based on skyrmions can be realized.

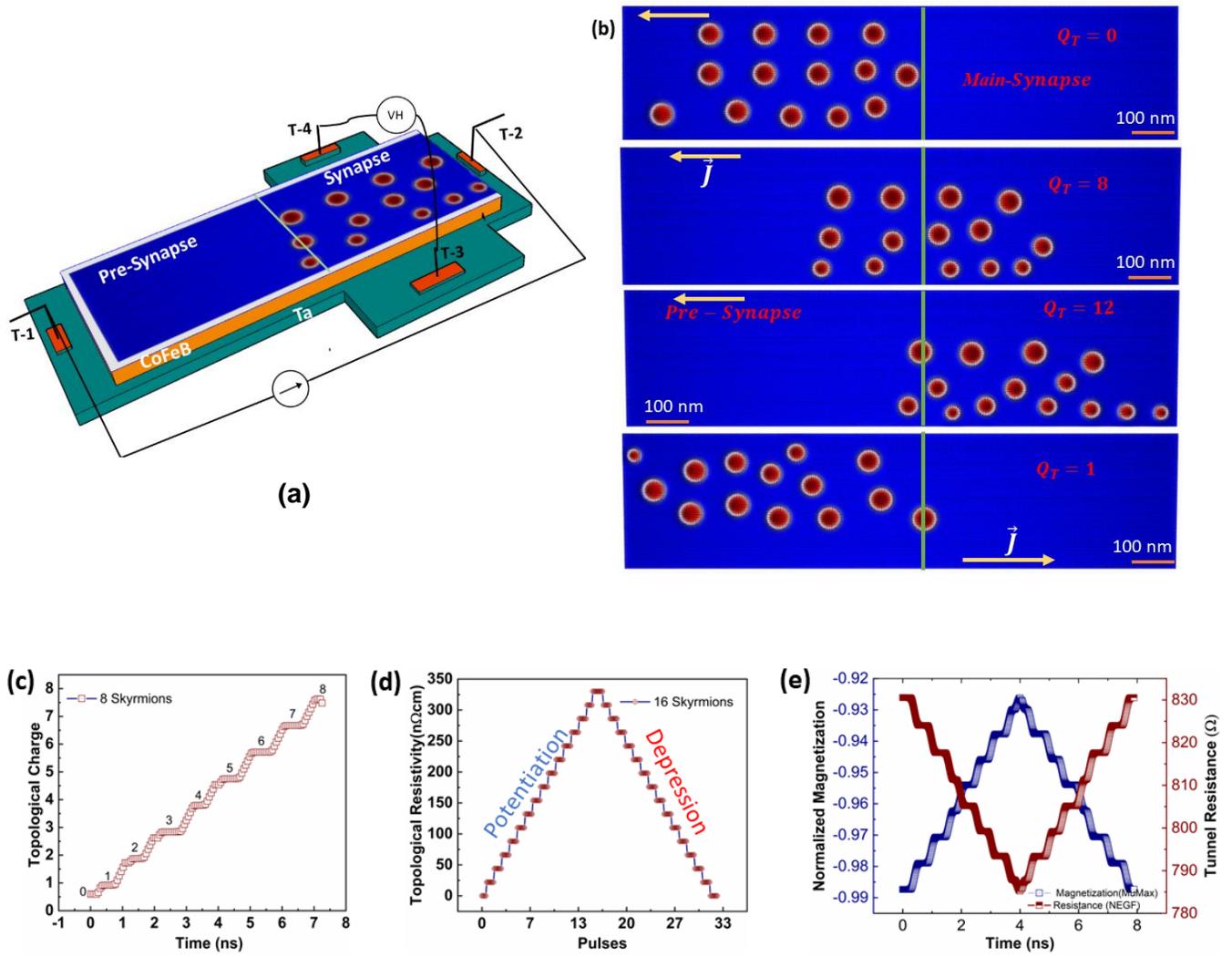

*Fig. 7 (a) Topological resistivity–based skyrmion synapse; current is applied across T-1 and T-2. (b) Magnetic texture of the synapse showing the skyrmion motion from the pre-synapse region into the main-synapse region and vice-versa; the corresponding topological resistivity change is measured across T-3 and T-4. (c) Evolution of the topological charge in the presence of continuous current ($8 \times 10^{11} A/m^2$). (d) Discrete topological resistivity of the device (measured across T-3 and T-4) for 16 skyrmions; by moving the skyrmions into/out of the main synapse, we achieve potentiation and depression. (e) Tunnel magneto-resistance in the MTJ configuration.*

*Synaptic weight and neural network configuration*

The synapse value is configured using a differential pair of skyrmion devices. To measure the synaptic weight, we employ two skyrmion devices to subtract the values and obtain the corresponding positive and negative weights [32]. The target synapse values are determined by the following equation:

$$G_{i,j}^{target} = G_{i,j}^{+} - G_{i,j}^{-} = k(|w_{i,j}^{+} - w_{i,j}^{-}|) = kw_{i,j}^{target},$$

where $G_{i,j}^{target}$ indicates the target conductance which can have positive or negative depending upon the difference between. $G_{i,j}^{+}$ and $G_{i,j}^{-}$, which indicate the device conductance receiving positive and negative voltage stresses at the site $(i,j)$th in the array.. Similarly, $w_{i,j}^{target}$ is the target weight obtained from the devices, and $k$ is the coefficient that relates the weights to the conductance. Fig. 8(a) shows the example of the synaptic weights, which are directly obtained from eight discrete skyrmion states. Fig. 9(a) shows the schematic circuit diagram considered for the vector-matrix multiplication operations with differential skyrmion device pairs.

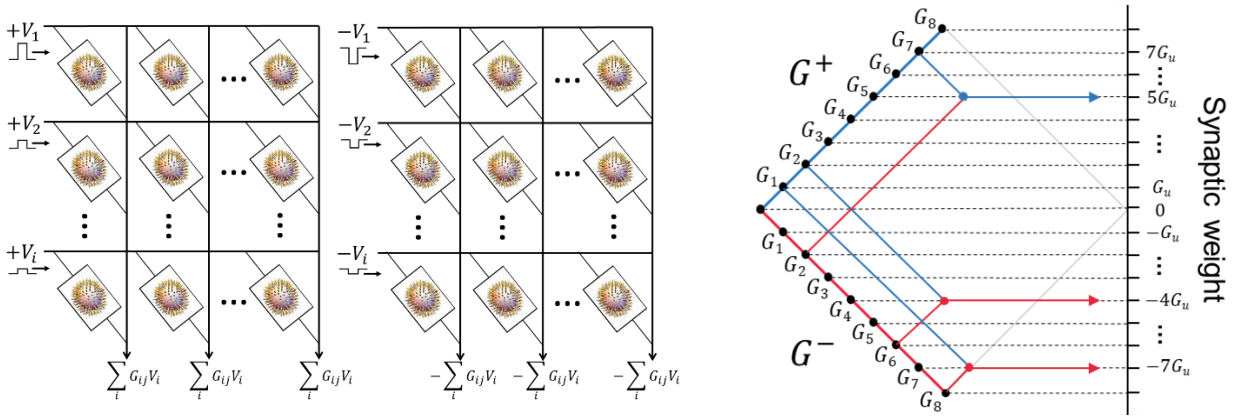

*Fig.8(a): Schematic of the circuit diagram comprising skyrmion-based synapses. (b) G diamond plot describing the method through which two skyrmion devices map to the synaptic weights, where each skyrmion device takes one of the eight conductance values from G1 to G8.*

Through Kirchhoff's law, the weighted current sum can simply be calculated as the result of matrix multiplication, which increases the computing speed and decreases the power consumption within the in-memory computing architecture.

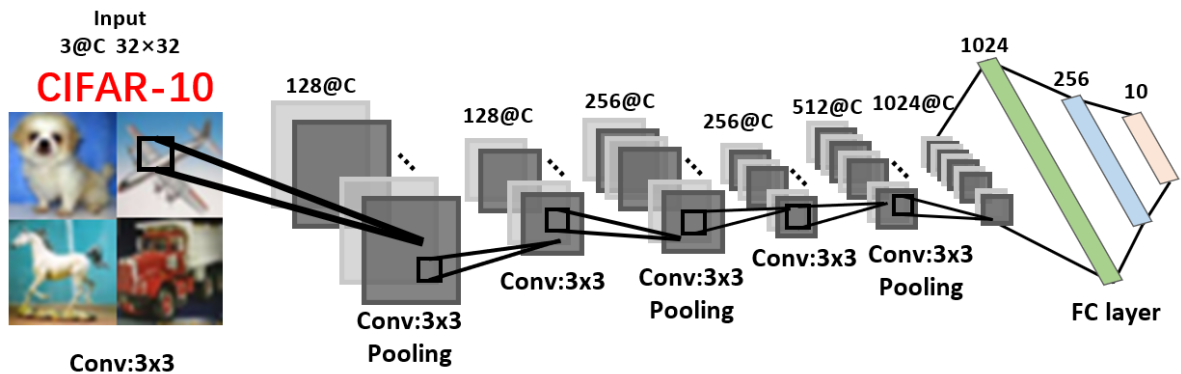

(a)

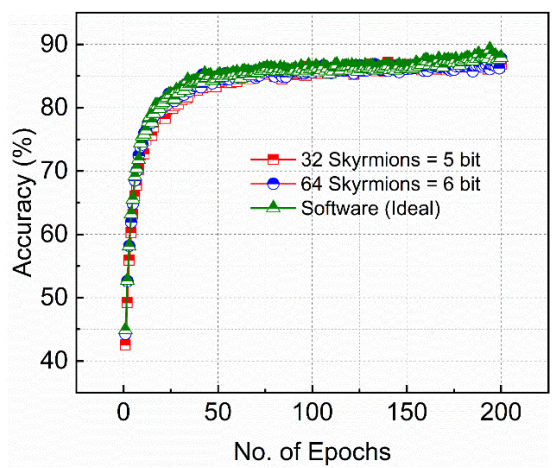

(b)

*Fig. 9(a): Neural network structure for the CNN model (a variant of VGG-NET) comprising six convolutional layers (CONV), three max-pooling layers (Pooling), and three fully connected layers (FC). (b) System-level performance in terms of the recognition accuracy.*

To measure the learning capability of our skyrmion device, we conducted an image-recognition task on CIFAR-10 Dataset with a nine-layer CNN, which is a brief variant of VGG-NET [33]. The network structure is shown in Fig. 9(a). Our network architecture comprises six convolutional layers for feature extraction and three fully connected layers for image classification. After every two convolutional layers, we adopt a max-pooling layer behind to subsample and aggregate the feature. In the simulation, the CNN weights are directly used to obtain the synaptic values from the skyrmion device. The topological resistivity of the skyrmion device has the advantages of excellent uniformity and linearity, which are beneficial for the implementation of the QCNN. For example, our skyrmion device has exhibited suitable properties on both LTP and LTD with 64 and 32 states. We can easily implement 5-bit and 4-bit quantized parameter networks separately with our device synapse.

The simulation results show that QCNN implemented with a skyrmion-based synaptic device achieves results comparable to those of software-based CNN algorithms. The simulation results are illustrated in Fig. (9b), we implement 5- and 6-bit synaptic weights with 32 and 64 skyrmions in the active region respectively to achieve the recognition accuracy around 87.76% and 86.81%, which is slightly lower than the 32-bit floating-point (FP32) arithmetic by software default. The experimental results demonstrate the learning capability of the device to achieve competitive accuracy in image recognition and highlight its applicability as a synaptic device for neuromorphic computing systems.

## V. CONCLUSIONS

In this study, we investigated the creation, stability, and controllability of skyrmions using experimental and simulation techniques to understand their applications in data storage and computing. We then analyze multiple aspects of the skyrmion stability, size, and density modulation under the influences of MF and anisotropy. Detailed insights into the transition from the labyrinth domains to skyrmions, along with the topological charge evolution, are obtained. Subsequently, an analytical model is developed to demonstrate the relationship between the skyrmion size and anisotropy, which helps in realizing VCMA-controlled synapses and neurons. We then analyze the influence of the DMI and anisotropy on the skyrmion size and density for device-parameter optimization in multiple skyrmion applications. Our results, in particular, contribute to the understanding of skyrmion voltage switching for data storage and unconventional computing applications. Therefore, we propose a skyrmion-based synaptic device based on our results and demonstrate the SOT-controlled skyrmion device with discrete topological resistance states. The discrete topological resistance of skyrmion device shows the inherent advantage of high linearity and uniformity, which makes it a suitable candidate for weight implementation of QCNN. The neural network is trained and tested on CIFAR-10 dataset, where we adopt the devices as synapses to achieve a competitive recognition accuracy against the software-based weights.